# Simultaneous 3D Construction and Imaging of Plant Cells Using Plasmonic Nanoprobe Assisted Multimodal Nonlinear Optical Microscopy

*Kun Liu,[1] Yutian Lei,[2] Dawei Li[1],\**

[1] School of Optoelectronic Engineering and Instrumentation Science, Dalian University of Technology, Dalian, Liaoning 116024, China

[2] Department of Civil Engineering, University of Nebraska-Lincoln, Lincoln, NE 68588, United States

\* Address correspondence to: dwli@dlut.edu.cn

**Abstract**: Nonlinear optical (NLO) imaging has emerged as a promising plant cell imaging technique due to its large optical penetration, inherent 3D spatial resolution, and reduced photodamage, meanwhile exogenous nanoprobes are usually needed for non-signal target cell analysis. Here, we report in-vivo, simultaneous 3D labeling and imaging of potato cell structures using plasmonic nanoprobe-assisted multimodal NLO microscopy. Experimental results show that the complete cell structure could be imaged by the combination of second-harmonic generation (SHG) and two-photon luminescence (TPL) when noble metal silver or gold ions are added. In contrast, without noble metal ion solution, no NLO signals from the cell wall could be acquired. The mechanism can be attributed to noble metal nanoprobes with strong nonlinear optical responses formed along the cell walls via a femtosecond laser scan. During the SHG-TPL imaging process, noble metal ions that cross the cell wall could be rapidly reduced to plasmonic nanoparticles by fs laser and selectively anchored onto both sides of the cell wall, thereby leading to simultaneous 3D labeling and imaging of potato cells. Compared with traditional labeling technique that needs in-vitro nanoprobe fabrication and cell labeling, our approach allows for one-step, in-vivo labeling of plant cells, thus providing a rapid, cost-effective way for cellular structure construction and imaging.

1. Introduction

Nonlinear optics (NLO), including multiphoton luminescence, second-harmonic generation (SHG), third-harmonic generation (THG), four-wave mixing (FWM) and coherent anti-stokes Raman scattering (CARS), has been widely applied in the biological fields since its inception.[1-9] Unlike linear optical microscopy, NLO microscopy offers inherent three-dimensional (3D) spatial resolution, relatively large optical penetration into tissues, and reduced photodamage,[10-11] which allows for viewing cells that are living and functioning. In most cases, more than one NLO microscopies are needed for biological studies since different organic components might have different NLO responses. The multimodal NLO imaging system has offered an approach to compare multi-images from different quantum processes, providing rich information for the study of various biological systems.[12-14]

On the other hand, although many ordered structures in cells and tissues can produce inherent NLO signal, exogenous probes are frequently used to increase the contrast or to selectively image non-signal targets. For example, exogenous SHG probes, including SHG dye or SHG nanoprobe, could be employed to extend its application to nonsignal target such as lipid bilayer in cells.[15] Although dyes have been extensively investigated for using as SHG probes, they suffer from photobleaching. Therefore, an active research area of interest for researchers is the development of novel nanoprobes that are less prone to bleaching and blinking while maintaining strong NLO responses.[16-19]

Compared with normal dye probes, noble metal nanoprobes are photostable, non-photobleaching, and biocompatible. Considerable research has proved that noble metal nanostructures such as silver and gold nanoparticles (NPs) exhibit various NLO properties, including two-photon luminescence (TPL),[20] SHG,[21-22] THG,[23] and FWM,[24] thus enabling them to be used as nanoprobes for labelling and imaging.[10]

Traditionally, noble metal nanoprobes must be fabricated externally and then transported inside the cell through functional groups or bonded antibody. Moreover, the size of the noble metal NPs must be well controlled to pass through the cell membranes,[25] which makes the labelling process complicated, challenging, and laborious.

As we know, plant cell walls are mostly composed of cellulose, which are invisible with conventional NLO imaging techniques. In this work, we report for the first time an innovative approach for in-vivo and simultaneous 3D labeling and imaging of complete potato cell structure using plasmonic nanoprobe assisted multimodal NLO microscopy. Instead of fabricating noble metal NPs outside the cell and then transporting exogenous NPs into the cells, we added noble metal ion solution onto the cells directly during the femtosecond (fs) laser imaging process. The obtained results show that noble metal ions could cross the cell membrane easily by osmotic pressure and are rapidly reduced to plasmonic NPs in-situ *via* a fs laser scan. The as-grown noble metal NPs are found to exhibit strong second-order (TPL, SHG) and third-order (FWM) NLO responses. More interestingly, these noble metal nanoprobes are selectively anchorited onto the cell walls. Through this approach, the whole structure of potato cells could be simultaneously labelled and imaged. Compared with traditional labelling techniques, our approach offers a simple and facile way for in-vivo constructing and 3D imaging of plant cells.

## 2. Results and Discussion

### 2.1 Characterization of potato cells using common imaging techniques

Potato cells mainly contain starch granules and cell walls, which makes it a simple plant cell structure for our study. Actually, starch granules exist in most of the plant cells and serve as the carbon storage, which are important starting materials for various industrial applications including biotechnology and biofuel production. The plant cell wall, which is composed of polysaccharides cellulose, hemicellulose, and pectin, acts as a semi-permeable layer that precisely permits the entry of smaller particles and inhibits the larger ones. To find out the complete structural information of potato cells, common imaging characterizations such as scanning electron microscopy (SEM) and Raman spectroscopy were first performed. To acquire the SEM imaging, potato samples need to be fully dehydrated by chemical fixation, due to the high vacuum chamber. Thus, the natural structure of potato cells might be altered and structure artifact might be caused. Figure 1a-1b show the typical SEM images of potato cells. As shown in Figure 1b, the starch granules and cell wall are clearly observed. However, the cell wall structure has been denaturized and even damaged due to the drying process.

Raman spectroscopy as one of the non-invasive techniques yields highly compound specific information for chemical analysis and has great potential for direct imaging. To reveal the intrinsic structure of potato cells, we carried out Raman measurement on the fresh potato. Figure 1c compares the Raman spectra taken on starch granules and cell wall of a fresh potato cell. The starch granules show significant Raman peaks in wavenumber ranging from 300-1800 $cm^{-1}$. However, no obvious Raman peaks for cell wall are observed in the same range of interest. Figure 1d displays the Raman mapping of the same potato sample in Figure 1c insert plotted using peak intensity at 477 $cm^{-1}$, further confirming that the structure of cell wall cannot be imaged but only starch granules are visible. In addition, only 2D imaging can be provided by Raman measurement, and it takes a long time (~2 hours) for mapping data acquisition. In comparison, the home-made multimodal nonlinear microscopy is capable to provide in-vivo 3D structure information of biological samples with a scan depth of up to one millimeter in a few minutes.

## 2.2 Nonlinear optical properties of potato cells

To investigate the feasibility of imaging of potato cell structures using multimodal nonlinear optical microscopy, we first measured the nonlinear optical spectra of potato cells. As shown in Figure 2a, two fs laser beams are used, one is a pump laser ($\omega_{pump}$) with wavelength of 800 nm and the other is a probe laser ($\omega_{probe}$) with broad wavelength between 850 and 1100 nm. The pump-probe laser beams are linearly polarized and collinearly focused onto the fresh potato samples without any pretreatment. Figure 2b shows the nonlinear optical spectra obtained from the potato starch granules (solid curve) and the potato cell walls (short dashed curve). It is obvious that the resulting spectrum of starch granules consist of four significant signal peaks: a sharp peak at a wavelength of ~400 nm with $\Delta\lambda$ = 7 nm, a strong peak at a wavelength of ~450 nm with $\Delta\lambda$ = 21 nm, a broad weak peak at a wavelength of 500-576 nm, and a broad peak at a wavelength of ~655 nm with $\Delta\lambda$ = 12 nm, labeled as "Peak A", "Peak B", "Peak C", and "Peak D", respectively. Since $\omega_A = 2\omega_{pump}$, the sharp peak at ~400 nm is the SHG emission from starch granules, which is consistent with the previous reports.[11, 26] "Peak B" is originated from the sum-frequency generation (SFG) process because this emission peak is observed at frequency $\omega_B = \omega_{pump} + \omega_{probe}$. Similarly, "Peak D" is attributed to the FWM process since the peak position corresponds to the frequency $\omega_D = 2\omega_{pump} - \omega_{probe}$.[27] The relatively weak broad peak at ~550 nm is originated from the TPL process by interaction with both laser beams. The intrinsic TPL signal of starch is so weak that it is almost invisible in comparison with the strong SHG. To summarize, under excitation by pump-probe laser beams, both second-order (SHG, SFG, TPL) and third-order (FWM) nonlinear optical behaviors have been simultaneously detected from potato starch granules. However, no significant NLO signals from potato cell walls is observed.

Next, we used nonlinear optical microscopy to perform simultaneous SHG-TPL-FWM optical imaging of potato cells. Figure 3a-3d show the SHG, TPL, FWM, and overlaid SHG-TPL-FWM images of a potato cell, respectively. The cell walls are marked by white dotted lines. It is found that only potato starch granules are visible, while no structure information on potato cell walls is observed. This observation is in accordance with the NLO spectra acquired in Figure 2b. Another interesting observation is that the SHG, TPL, and FWM images can provide different but complementary information for each other. Figure 3e-3h displays the corresponding SHG, TPL, FWM images with bigger magnification in Figure 3a-3c. It can be seen from Figure 3e and 3f that the distribution of TPL signal at potato starch grains is complementary to that of SHG. The similar phenomenon has also been observed for collagen whose SHG and TPL signals highly depend on their space arrangement.[28] In comparison, FWM imaging (Figure 3g) has relatively weaker contrast than that of TPL and SHG, with the strongest FWM signal distributed at the edge of the starch granules.

From above analyses, it can be concluded that, to realize imaging of complete potato cells using multimodal nonlinear optical microscopy, an effective labeling approach for cell walls must be developed. In next section, we investigate the possibility of fs laser assisted growth of noble metal nanoparticles (NPs) as nanoprobes for cell structure labeling and imaging.

**2.3 Nonlinear optical properties of fs laser induced noble metal NPs**

Previous studies have shown that noble metal ions, such as silver ions ($Ag^+$), could be reduced to silver NPs under fs laser scan due to multi-photon absorption or thermal effect.[29-30] To monitor noble metal NP formation process and its nonlinear optical properties, we performed in-situ, real-time nonlinear optical spectral measurement in silver ion solution under a focused fs laser irradiation. Figure 4a shows the nonlinear optical spectra acquired on fused silica substrate in silver ion solution as a function of fs laser irradiation times. Each spectrum has an accumulation time of 0.2 s. At the beginning, only slight background signal at around 665 nm is observed since few or no silver NPs are formed. As fs laser irradiation time goes on, significant SHG peak signal centered at ~400 nm, TPL signal with a broad peak at 475-625 nm, and broad FWM peak signal at~655 nm has been observed. Figure 4b reveals the evolution of nonlinear optical signals of SHG, TPL, and FWM extracted from Figure 4a. With increasing laser irradiation time, the SHG signal (lower panel, Figure 4b) is increased quickly, reaches a maximum value at the time of ~10 s, reflecting that silver nanoparticle structures have been formed on the substrate. However, with further increasing laser irradiation time, the SHG signal is decreased by an order of magnitude, which suggests that silver NPs have aggregated into cluster structures, as evidenced by SEM characterization (Figure 4a insert). For TPL (Figure 4b, middle panel), the signal is slowly increased at the beginning, reaches the maximum at around 15 s, and then maintains a stable and strong value with further laser irradiation. The FWM signal (Figure 3b, upper panel) has the similar trend as that of TPL, which takes around 14 s to reach its maximum value. These analyses indicate that within a short time of fs laser irradiation (just a few seconds), silver NPs with strong nonlinear responses have been be formed. In addition, the strong nonlinear optical responses have been detected in gold NPs (Supporting Information), suggesting that fs laser induced both silver and gold nanostructures are suitable to be used as nanoprobes.

To demonstrate the capability of fs laser induced noble metal NPs for real-time imaging, we performed 2D imaging acquisition by involving a square scan of fs laser in noble metal ion solution. Figure 4c shows the TPL image of a fused silica substrate after fs laser scan in silver ion solution, where the laser scanned area exhibits uniform and strong nonlinear optical signal. It reveals that the patterned silver NPs with uniform size distribution have been formed, as confirmed by SEM imaging characterization (Figure 4d). Moreover, this experiment demonstrates that in-situ silver ion reduction and NP formation could be realized during the nonlinear imaging process, further confirming that fs laser scan induced noble metal NPs can be used as nonlinear optical nanoprobes for nonsignal target imaging.

**2.4 3D construction and imaging of potato cells via fs laser scan in noble metal ion solution**

Figure 5a shows the schematic of our proposed approach for one-step 3D cell structural construction and simultaneous potato cell imaging via fs laser scan in noble metal ion ($Ag^+$ or $Au^{3+}$) solution. It is considered that, during the nonlinear optical imaging process, noble metal ions that cross the cell wall (middle panel) could be rapidly reduced to NPs by fs laser and selectively anchored onto both sides of the cell wall (lower panel), thus leading to simultaneous 3D labeling and imaging of complete potato cell structures. To test this idea, two drops (~5 mL) of silver ion solution were added on a thin slice of a freshly cut potato sample (Figure 5b). Then, multimodal 2D (*x-y* plane) nonlinear optical imaging of potato cells with a scan depth (*z*) of 150 μm was simultaneously acquired. The details of the revolution of cell wall structure construction with and without adding silver ion solution was also recorded (Supporting Information). After that, 3D nonlinear optical images of potato cells can be obtained by projecting all the 2D images in the *z* direction (Figure 5c-5e).

For comparison, we first collected multimodal 3D nonlinear optical images of potato cells without adding silver ion solution (Figure 5c), where all the images including TPL (left), SHG (middle), and overlaid TPL-SHG (right) only show information of starch granules. We also note that the SHG image offers a better signal to noise ratio and stronger contrast than that of TPL. This is in accordance with our observed results in Figure 2b that starch has much stronger SHG than TPL. With silver ion solution being added (Figure 5d), both TPL and SHG images show the structure information of cell walls, confirming that silver NPs have been formed during imaging process and uniformly anchored onto the cell walls. A careful observation further reveals that the TPL image displays the cell wall structure with sharp contrast, while the corresponding SHG image shows both starch grains and cell wall structure, but with cell wall having a much weaker contrast than that in TPL. When merging SHG with TPL together, the whole potato cell structure including starch granules and cell walls could be clearly imaged, as shown in overlaid TPL-SHG image in Figure 5d.

To investigate the influence of fs laser scan time on the cell wall structure labelling, the same potato sample in Figure 5b was imaged for a second time (Supporting Information). Figure 5e shows the 3D nonlinear optical imaging acquired with the second time of fs laser scan in silver ion solution. In comparison with the first scan (Figure 5d), there is no obvious enhancement in SHG or TPL signal intensity or contract. Thus, a single laser scan is enough for silver ion reduction, NP formation, labelling and imaging of cell structures. Besides, similar results have also been observed by performing identical experiments in gold ion solution (Supporting Information), further supporting our idea that fs laser induced noble metal NPs could be used as very good nonlinear optical nanoprobes for in-situ and simultaneous plant cell labeling and imaging. Thus, our 3D labelling and imaging approach is one-step, simple and cost-effective by just imaging the cells in noble metal ion solution without any other treatments.

## 3. Conclusions

In summary, we have developed a simple, one-step approach for 3D plant cell structure labeling and imaging by using noble metal ion assisted multimodal nonlinear optical microscopy. Instead of fabricating noble metal nanoprobes outside and then transporting into cell for labeling/imaging, we realized fs laser assisted noble metal NP growth and in-vivo labeling during the 3D imaging process in noble metal ion solution. We found that fs laser induced silver NPs exhibit various excellent nonlinear optical responses including TPL, SHG and FWM. In addition, these silver NPs can be selectively anchored onto the cell walls, thus enabling the simultaneous 3D labelling and nonlinear optical imaging of complete potato cell structures. The similar results have also been observed when substituting silver ions with gold ions. Thus, our method offers a new avenue for facile and in-vivo 3D labelling/imaging of plant cells and other cell structures, which is very useful for future biological applications.

## 4. Methods

**Reagents and materials**. Silver nitrate (purity > 99 %), gold (III) chloride trihydrate (> 99.9% trace metal basis) were purchased from Sigma-Aldrich (St. Louis, MO, USA) and used without further purification. Deionized water with electrical resistivity of 18 MΩ-cm was used for preparing silver ion and gold ion solutions with a concentration of 0.3 mM and 3.7 mM, respectively. Fresh potatoes were cut into thin slices and fixed onto glass slides.

**Multimodal nonlinear optical imaging of potato cells**. The nonlinear optical properties of as-prepared and noble metal NPs-decorated potato cell samples were investigated using home-made multiphoton nonlinear optical microscopy system. A commercial Ti:Sapphire fs laser with a center wavelength of 800 nm was used as the laser source. Combing with a supercontinuum generator, two incident laser beams with tunable powers could be generated. One is called pump laser beam, which is formed by introducing 800 nm laser beam through an attenuator and a delay line. The other is called probe laser, which is formed by using a 500 mW laser beam to generate the supercontinuum which is then filtered through a long-pass filter. The pump-probe laser beams were then collinearly focused onto the sample surfaces. A spectrometer was used for the nonlinear optical spectra measurement. The nonlinear optical imaging signals were collected using PMTs. By applying a 390/40 nm emission filter, a 495-540 nm band-pass filter, and a 647/57 nm band-pass filter before PMTs, three different nonlinear optical images could be simultaneous acquired.

**Other characterizations**. Raman analyses were performed at room temperature in a micro-Raman spectrometer (Renishaw InVia plus, Renishaw, Gloucetershire, U.K.). Raman spectra and mapping were collected through a 50× objective lens with an exposure time of 10 s and 1 s, respectively, at each position. Morphological characterization of potato cells and noble metal particles was carried out using a field emission scanning electron microscopy (SEM, JEOL JSM-7600F).

**Acknowledgements**

This work was supported by the National Natural Science Foundation of China (Grant No. 12274051), the "Chunhui Project" Cooperative Research Project of Ministry of Education (Grant No.

HZKY20220423), and the Fundamental Research Funds for the Central Universities (Grant No. DUT21RC(3)032).

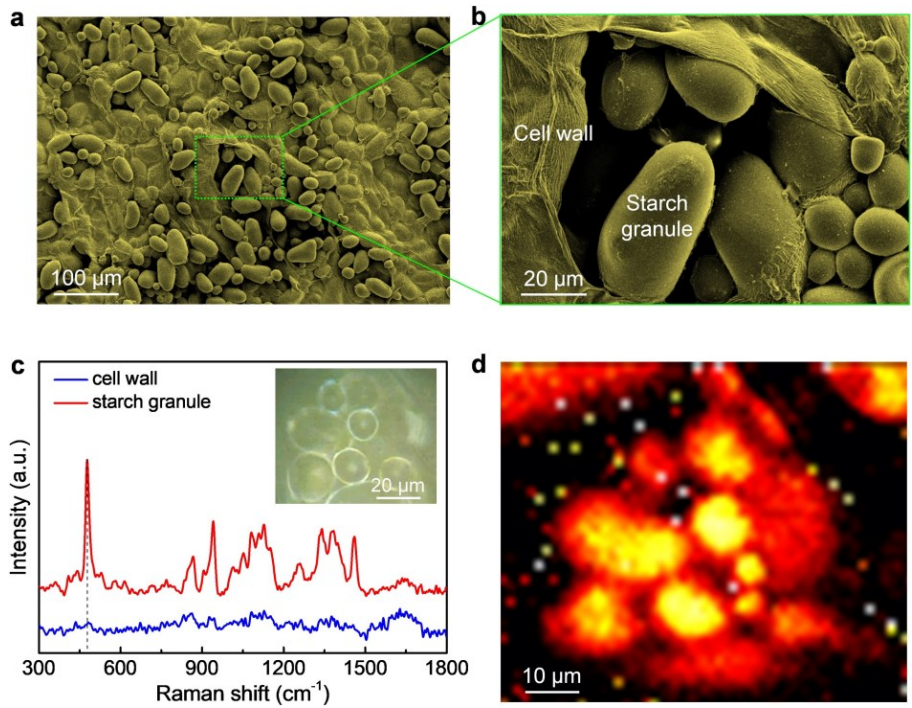

**Figure 1.** Normal imaging techniques for potato cells characterization. (a) and (b) SEM images of potato cells with different magnifications. (c) Raman spectra of the potato cell wall (red) and starch granule (blue). Inset: optical image of the sample area for Raman measurement. (d) Raman mapping of potato cell sample in (c) using the peak intensity at 477 cm$^{-1}$.

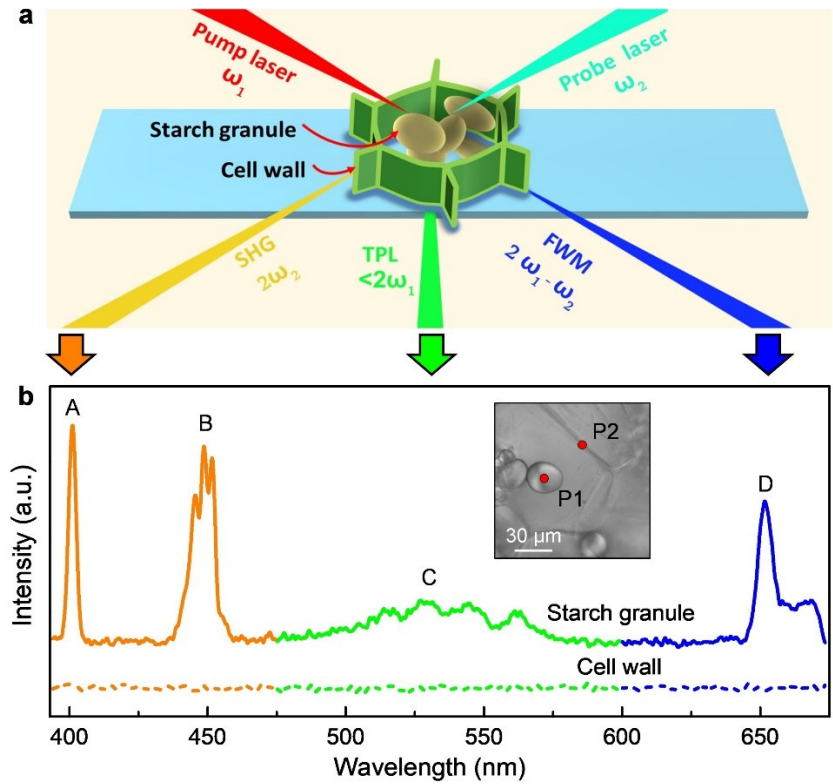

**Figure 2.** Multiphoton nonlinear optical spectral analyses of potato cells. (a) The schematic of interaction of pump-probe laser beams with a potato cell, where the second-order (SHG, TPL) and third-order (FWM) nonlinear optical signals are simultaneously detected. (b) Nonlinear optical spectra of starch granule (P1) and cell wall (P2) generated by pump-probe laser beams. The measurement points are labeled in the optical image (inset).

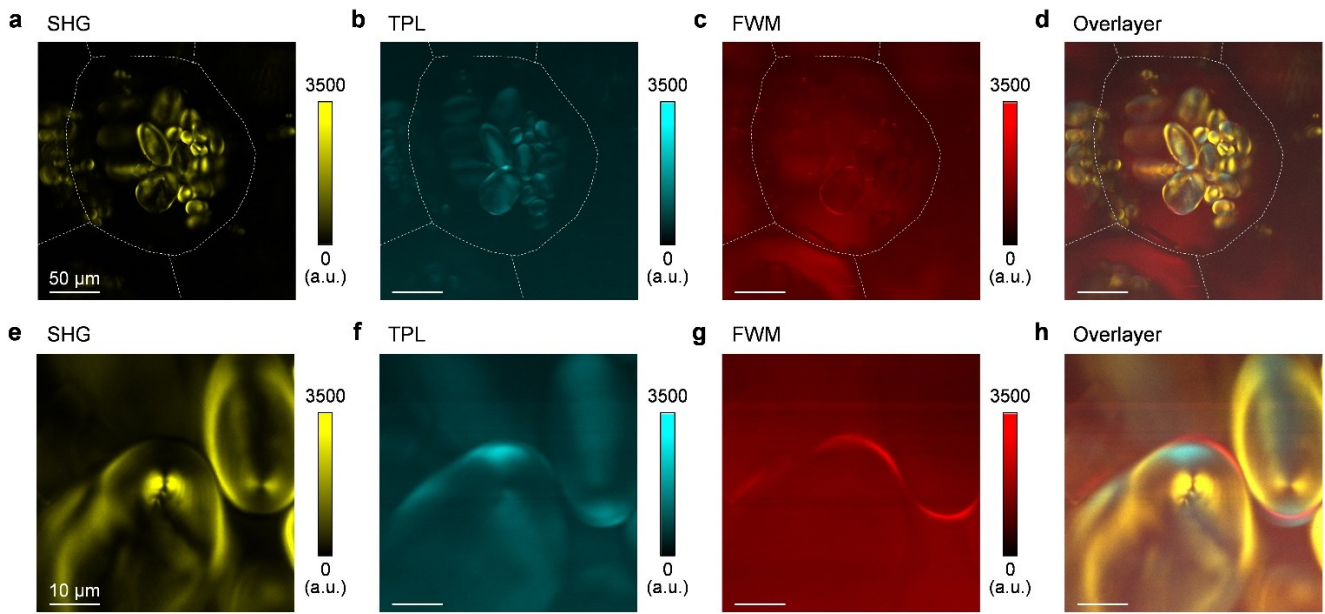

**Figure 3.** Simultaneous and multimodal nonlinear optical imaging of a potato cell: (a) SHG, (b) TPL, (c) FWM, and (d) overlaid SHG-TPL-FWM. The dashed lines in (a-d) indicate the cell wall of the potato. (e-f) The magnified (e) SHG, (f) TPL, (g) FWM, and (h) overlaid SHG-TPL-FWM images of the starch granule in (a-d). The scale bars are 50 μm for (a-d) and 10 μm for (e-h).

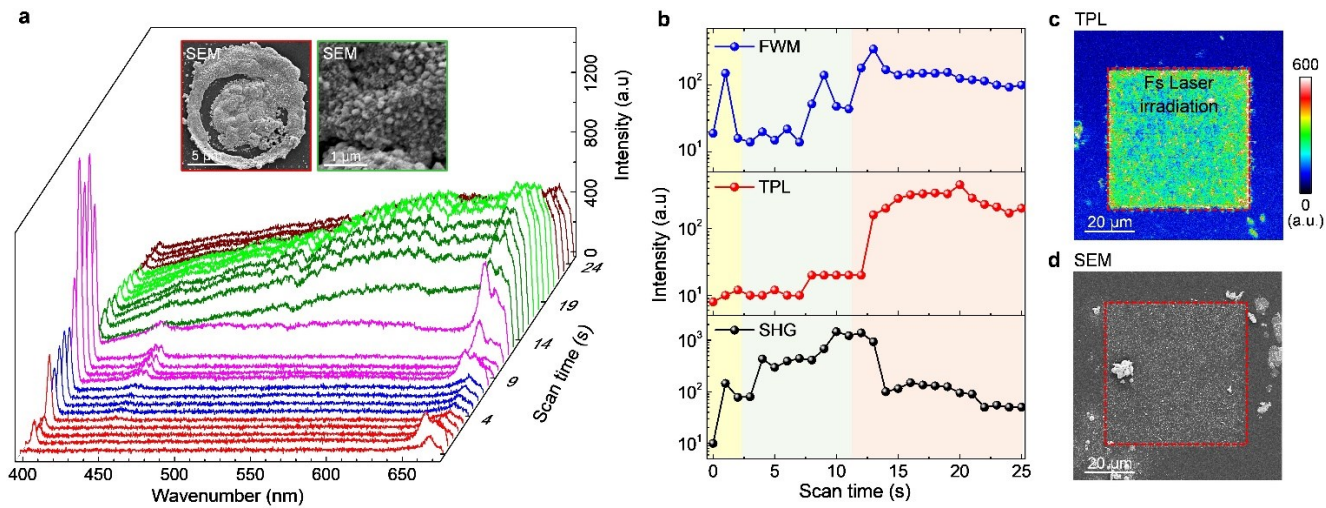

**Figure 4.** Femtosecond laser induced silver NPs growth and its nonlinear optical responses. (a) Nonlinear optical spectra acquired on fused silica substrate in silver ion solution under a focused fs laser irradiation with different accumulation time. (b) The peak intensity of SHG (lower), TPL (middle), and FWM (upper) extracted from (a) as a function of fs laser irradiation time. Inset in (a): SEM images of silver nanoparticle aggregates formed with a long-time fs laser irradiation. (c) TPL mapping of a fused silica substrate in silver ion solution under fs laser scan in a square pattern (yellow dashed line), (d) with the corresponding SEM image. As expected, patterned silver NPs are grown.

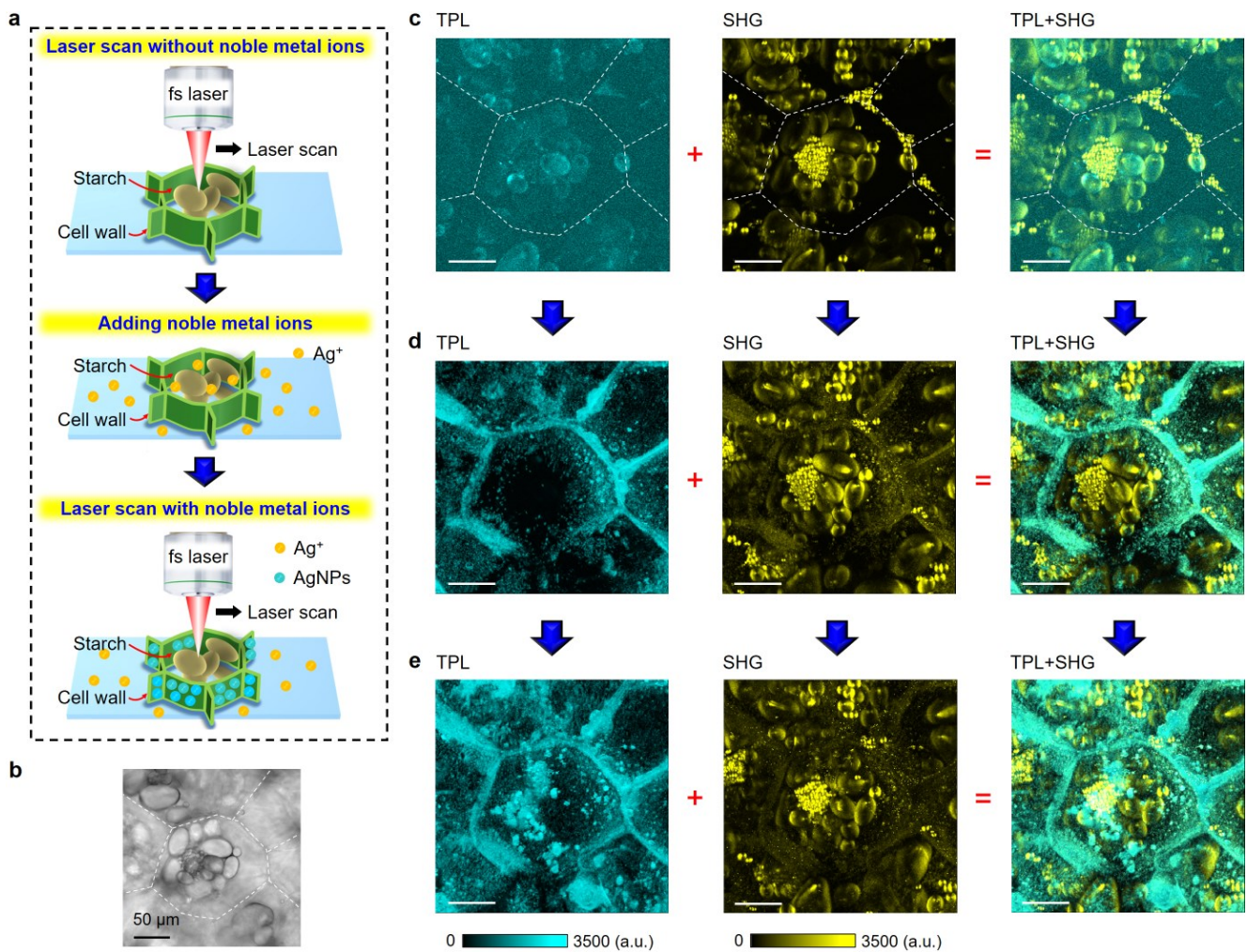

**Figure 5.** Simultaneous 3D construction and imaging of potato cells in silver ion solution using multimodal nonlinear optical microscopy. (a) The schematic of fs laser scan induced imaging and construction of potato cells without (upper panel) and with (middle and lower panels) silver ion solution. (b) Optical image of a potato cell. (c) 2D Z-stacked TPL (left), SHG (middle), and overlaid TPL-SHG (right) images of the potato cell in (b) without adding silver ion solution. The dashed lines in (b-c) indicate the cell wall of the potato. (d) Corresponding TPL (left), SHG (middle), and overlaid TPL-SHG (right) images in (c), respectively, taken with fs laser scan only for one time in silver ion solution. (e) Corresponding TPL (left), SHG (middle), and overlaid TPL-SHG (right) images in (c) with fs laser scan for the second time in silver ion solution. The scale bar in (b-e) are 50 μm.

18